\documentclass{article}

\usepackage{natbib}

\def\R{\ifmmode{I\hskip -3pt R}
            \else{\hbox{$I\hskip -3pt R$}}\fi}

\begin{document}
\title{Clustering strategy and method selection}
\author{Christian Hennig, \\
Department of Statistical Science, University College London}

\maketitle

\begin{abstract} 
{\bf Note:} {\it This paper is a chapter in the forthcoming Handbook of Cluster 
Analysis, \cite{Handbook}. For definitions of basic clustering methods and
some further methodology, other chapters of the Handbook are referred to.
To read this version of the paper without the Handbook, some knowledge of
cluster analysis methodology is required.}

The aim of this chapter is to provide a framework for all the decisions that
are required when carrying out a cluster analysis in practice. A general 
attitude to clustering is outlined, which connects these decisions closely
to the clustering aims in a given application. From this point of view, the 
chapter then discusses aspects of data processing such as the choice of the
representation of the objects to be clustered, dissimilarity design, transformation
and standardization of variables. Regarding the choice of the clustering 
method, it is explored how different methods correspond to different 
clustering aims. Then an overview of benchmarking studies comparing different
clustering methods is given, as well as an outline of theoretical approaches
to characterize desiderata for clustering by axioms. Finally, aspects of 
cluster validation, i.e., the assessment of the quality of a clustering in a
given dataset, are discussed, including finding an appropriate number of clusters,
testing homogeneity,
internal and external cluster validation, assessing clustering stability and
data visualization.

% The current template proposes a structure for your handbook chapter (sections
% \ref{ch3.5sec:intro}-\ref{ch3.5sec:concl}). More guidelines fopr writing
% are given
% in sections \ref{ch3.5sec:gfw}-\ref{ch3.5sec:gn}.
% % Please give a short summary of your chapter in an abstract.
\end{abstract}

\tableofcontents

\section{Introduction} \label{ch6.6sec:intro}
{\bf Note:} {\it This paper is a chapter in the forthcoming Handbook of Cluster 
Analysis, \cite{Handbook}. For definitions of basic clustering methods and
some further methodology, other chapters of the Handbook are referred to.
To read this version of the paper without the Handbook, some knowledge of
cluster analysis methodology is required.}

In \cite{Handbook}, a large number of cluster analysis methods have 
been introduced, and in any situation in which a clustering is needed,
the user is faced with a potentially overwhelming number of options.
The current paper is about how the required choices can be made. \cite{Mil96} 
listed seven steps of a cluster analysis that require decisions, namely
\begin{enumerate}
\item choosing the objects to be clustered,
\item choosing the measurements/variables,
\item standardization of variables,
\item choosing a (dis-)similarity measure,
\item choosing a clustering method,
\item determining/deciding the number of clusters,
\item interpretation, testing, replication, cluster validation.
\end{enumerate}
I will treat all but the first one (general principles of 
sampling and experimental design apply), not sticking exactly to this order.
The chapter 
focuses on the general philosophy behind the required choices, what this means 
in practice, and on some areas of research. This has to be combined with
knowledge on clustering methods as given elsewhere in this volume. 
Some more discussion
of the above issues can be found in \cite{Mil96} and standard cluster 
analysis books such as \cite{JaDu88,KaRo90,Gordon99,EvLaLeSt11}. 

The point of view taken here, previously outlined in
\cite{HenLi13} and also shared by other authors (\cite{LuWiGu12}), 
is that there is no such thing as a universally 
"best clustering method". Different methods should be used for
different aims of clustering. The task of selecting a clustering method 
implies a proper understanding of the meaning of the data, the clustering
aim and the available methods, so that a suitable method can be matched
to what the application requires. Although many experienced experts in
the field, including the authors of the books cited above,
agree with this view, there is not much advice in the literature on how
the specific requirements of the application can be connected with the
available methods. Instead, cluster analysis methods have been often 
compared on simulated data or data with known classes, in order to find
a ``best'' one disregarding the research context. Such comparisons are of
some use, particularly because they reveal, in some cases, that methods may 
not be up for what they were supposed to do. Still, 
it would be more useful to have more specific information
about what kind of method is connected to what kind of clustering task,
defined by clustering aim, required cluster concept, and potential structure
in the data. 

The present chapter goes through the most essential steps of making the
necessary decisions for a cluster analysis. 
It starts in Section \ref{ch6.6sec:philo}
with a discussion of the background, relating the aims of clustering
to the cluster concepts that may be of interest in a specific situation. 
Section \ref{ch6.6:data} looks at the data to be clustered. 
Often it is useful to pre-process
the data before applying a clustering method, by defining new variables,
dissimilarity measures, transforming or selecting features. Such operations
have an often fundamental impact on the resulting clustering. Note that I
will use the term ``features'' to refer to the variables eventually used 
for clustering if a cluster analysis method for an
``objects times features''-matrix as input is applied, whereas the term 
``variables'' will be used in a more general sense for measurements 
characterizing the objects used
in the clustering process, potentially later to be used as clustering
features, or for computing dissimilarity measures or new variables.  

Section \ref{ch6.6:benchmark} is on comparing clustering methods. This 
encompasses the decision which method fits a certain clustering aim, 
measurement of the quality of clustering methods, benchmark simulation studies,
and some theoretical work on characterizing clusterings and clustering
methods.
In many cases, though, there may not be enough precise information about the
clustering aim and  cluster concepts of interest, so that the user may not
be able to pinpoint exactly what method is needed. Also, it may be discovered
that the clustering structure of the data may differ from what was expected 
in advance, and other methods than initially considered may look promising.
Section \ref{ch6.6:validation} 
is about evaluating and comparing outcomes of clustering 
methods, before the chapter is concluded.

\section{Clustering aims and cluster concepts}
\label{ch6.6sec:philo}
In various places in the literature it is noted that there is no generally
accepted definition of a cluster. This is not surprising, given the many
different aims for which clusterings are used. Here are some examples:
\begin{itemize}
\item delimitation of species of plants or animals in biology, 
\item medical classification of diseases,
\item discovery and segmentation of settlements and periods in archeology,
\item image segmentation and object recognition,
\item social stratification,
\item market segmentation,
\item efficient organization of data bases for search queries.
\end{itemize}
There are also quite general tasks for which clustering is applied in many
subject areas:
\begin{itemize}
\item exploratory data analysis looking for ``interesting patterns'' without 
prescribing any specific interpretation, potentially creating new research
questions and hypotheses,
\item information reduction and structuring of sets of entities from 
any subject 
area for simplification, more effective communication, or more effective 
access/action such as complexity reduction for further data analysis,
\item investigating the correspondence of a clustering in specific data
with other groupings or characteristics, either 
hypothesized
or derived from other data.  
\end{itemize} 
Depending on the application, it may differ a lot what is meant by a 
``cluster'', and this has strong implications for the methodological strategy.
Finding an appropriate clustering method means that the cluster definition and
methodology have to be adapted to the specific aim of clustering in
the application of interest.

A key distinction can be made between ``realist'' aims of clustering, 
concerning the discovery of some meaningful real structure corresponding to
the clusters, and ``constructive'' aims, where researchers intend to split up
the data into clusters for pragmatic reasons, regardless of whether there is 
some essential real difference between the resulting groups. This distinction 
can be roughly connected to the choice of clustering methodology. For example,
some clustering criteria such as $K$-means (\cite{Handbook}) produce 
homogeneous clusters in the sense that all observations are assigned to the 
closest centroid, and large distances within clusters are heavily penalized. 
This is useful for a number of constructive clustering aims. On the other hand, 
$K$-means does not pay much attention to whether or not the clusters are
clearly separated by gaps, and does not tolerate large variance and spread
of points within clusters, which can occur in clusters that correspond to 
real patterns (for example objects in images). 

However, the distinction between realist and constructive clustering aims is 
not as clear cut as it may seem at first sight. Categorization is a very basic 
human activity that is directly connected with the emergence of language. 
Whenever human beings speak of real patterns, this can only refer to 
categories that are aspects of human cognition and 
can be expressed in language, which can be seen as a pragmatic 
human construct (\cite{MeHaMiTh93} review cognitive theories of
categorization with a view to connecting them to inductive data analysis
including clustering). In a related manner, 
researchers with realist clustering aims  
should not hope that the data alone can reveal real structure; constructive 
impact of the researchers is needed to decide what counts as real.

The key issue in realist clustering is how the real 
structure the researchers are interested in is connected to the available 
data. This requires subject matter knowledge, but it also requires 
decisions by the researchers.   
``Real structure'' is often understood as the existence of an unobserved 
categorical variable the values of which define the ``true'' clusters. Such an
idea is behind the popular use of datasets with given true classes for 
benchmarking of cluster analysis methods. But neither can it be taken fur 
granted that the known categories are the only existing ones that could qualify
as ``real clusters'', nor do such categories necessarily correspond to data
analytic clusters. For example, male/female is certainly a meaningful 
categorization of human beings, but there may not even be a significant 
difference between men and women regarding the results of a certain attitude 
survey, let alone separated clusters corresponding to sex. Usually the objects
represented in the dataset can be partitioned into real categories in many
ways. Also, different cluster analysis methods will produce different 
clusterings, which may more or less well correspond to patterns that are 
real in potentially different ways. This means that in order to decide about
appropriate cluster analysis methodology, researchers need to think about
what data analytic characteristics the clusters they are aiming at are supposed
to have. I call this the ``cluster concept'' of interest in a specific study.

The real patterns of interest may be more or less closely connected to the 
available data. For example, in biological species delimitation, the concept of
a species is often defined in terms of interbreeding (there is some
controversy about the precise definition, see \cite{Hausdorf11}). 
But interbreeding 
patterns are not usually available as data. Species are nowadays usually 
delimited by use of genetic data, but in the past, and also occasionally in 
the present in an exploratory manner, species were seen as the source of a 
real grouping in phenotype data. In any case, the researchers need some 
idea about how true distinctions between species are connected to 
patterns in the data. Regarding genetic data, this means that knowledge needs
to be used about what kind of similarity arises from persistent genetic 
exchange inside a species, and what kind of separation arises between distinct
species. There may be subgroups of individuals in a species 
between which there is little 
actual interbreeding (because potential interbreeding suffices for 
species definition), for example between geographically separated groups, and 
consequently not as much genetic similarity as one would naively expect.
Furthermore there are various levels of classification in biology, such as
families and genii above and subspecies below the level of species, so that 
data analytic clusters may be found at several levels, and the researchers
may need to specify more precisely how much similarity within and separation
between clusters is required for finding species.

Such knowledge needs to be reflected in the cluster analysis method to be 
chosen. For example, species may be very heterogeneous regarding geographical
distribution and size, and therefore a clustering method that implicitly 
tends to bring forth clusters that are very homogeneous such as $K$-means or 
complete linkage is inappropriate. 

In some cases, the data are more directly connected to the cluster definition.
In species delimitation, there may be interbreeding data, in which 
case researchers can specify the requirements of a clustering more directly.
This may imply graph theoretic clustering methods and a specification of 
how much connectedness is required within clusters, although such decisions
can often not be made precise because of missing information arising from 
sampling of individuals, missing data etc. On the other hand, the connection
between the cluster definition and the data may be less close, as in the case
of phenotype data used for delimiting species, in which case the researchers
may not have strong information about how the clusters they are interested in 
are characterized in the data, and some speculation is needed in order to decide
what kind of clustering method may produce something useful.
 
In many situations different groupings can be interpreted as real, depending on
the focus of the researchers. Social classes for example can be defined in 
various ways. Marx made ownership of means of production the major 
defining characteristic of different classes, but social classes can also be defined 
by looking at patterns of communication and contact, or occupation, or 
education, or wealth, or by a mixture of these (\cite{HenLi13}). 
In this case, a major issue 
for data clustering is the selection of the appropriate variables and 
measurements, which implicitly defines what kinds of social classes can be 
found. 

The example of social stratification also illustrates that there is a gradual
transition rather than a clear cut between realist and constructive 
clustering aims. According to some views (such as the Marxist one) 
social classes are an essential and real characteristic of society, but according
to other views, in many societies there is no clear delimitation between
social classes that could justify to call these classes ``real'', despite 
the existence of real inequality. 
Social classes can then still be used as a convenient tool for
structuring the inequality in such societies. 

Regarding constructive clustering aims, it is obvious that researchers need
to decide about the desired ``cluster concept'', or in other words, about
the characteristics that their clusters should have. The discussion above
implies that this is also the case for realist clustering aims, for which
the required cluster concept needs to be derived from knowledge about the
nature of the real clusters, and from a decision of the researchers about
their focus of interest if (as is usually the case) the existence of
more than a unique real clusterings is conceivable. For constructive 
clustering, the required cluster concept needs to be connected to the 
practical use that is intended to be made of the clusters. 

Also where the primary clustering aim is constructive, realist aims may still
be of interest insofar as if indeed some real grouping structure is
clearly manifest in the data, many constructive aims will be served well by
having this structure reflected in the clustering. For example, market 
segmentation may be useful regardless of whether there are really 
meaningfully separated groups in the data, but it is relevant
to find them if they exist.

Here is a list of potential characteristics of clusters that may be desired,
and that can be checked using the available data. A number of these are 
related with the ``formal categorization principles'' listed in Section 
14.2.2.1 of \cite{MeHaMiTh93}.
\begin{enumerate}
\item Within-cluster dissimilarities should be small. 
\item Between-cluster dissimilarities should be large.
\item Clusters should be fitted well by certain homogeneous 
probability models such as the Gaussian or a uniform distribution on 
a convex set, or, if appropriate, 
by linear, time series or spatial process models.  
\item Members of a cluster should be well represented by its centroid.
\item The dissimilarity matrix of the data should be well represented by the
clustering (i.e., by the ultrametric induced by a dendrogram, or by defining
a binary metric ``in same cluster/in different clusters''). 
\item Clusters should be stable.
\item Clusters should correspond to connected areas in data space with high 
density. 
\item The areas in data space corresponding to clusters should have certain
characteristics (such as being convex or linear).
\item It should be possible to characterize the clusters using a small number
of variables.
\item Clusters should correspond well to an externally given partition or 
values of one or more variables that were not used for computing the 
clustering. 
\item Features should be approximately independent within clusters.
\item All clusters should have roughly the same size.
\item The number of clusters should be low.
\end{enumerate}
When trying to measure these characteristics, they have to be made more precise,
and in some cases it matters a lot how exactly they are defined. Take no. 1,
for example. This may mean that all within-cluster dissimilarities 
should be small
without exception (i.e., the maximum should be small, as required by complete 
linkage hierarchical clustering), or their average, or a high quantile of 
them. These requirements may look similar at first sight but are very 
different regarding the integration of outliers in clusters. Having small 
within-cluster dissimilarities may emphasize gaps by looking at the smallest
dissimilarities between each two clusters, or it may rather mean that the 
central areas of the clusters are well distributed in data space. 
 As another example, stability can refer to sampling other data from the same
population, to adding ``noise'', or to comparing results from different 
clustering algorithms. 

Some of these characteristics are in conflict with others in some
datasets. Connected areas with high density may include very large distances,
and may have undesired (e.g., non-convex or nonlinear) shapes. Representing 
objects by centroids may bring forth some clusters with little or no 
gap between them. Having clusters of roughly equal size forces
outliers to be integrated in distant clusters, which produces 
large within-cluster dissimilarities.

Deciding about such characteristics is the key to linking the clustering 
aim to an appropriate 
clustering method. For example, if a database of images should be clustered 
so that users can be shown a single image to represent a cluster, no. 7 is 
most important. Useful market segments need to be addressed by 
non-statisticians and should therefore normally be represented by few
variables, on which dissimilarities between members should be low. Similar
considerations can be made for realist clustering aims, see above.

For choosing a clustering method, it is then necessary to know how they 
correspond to the required characteristics. Some methods optimize certain
characteristics directly (such as $K$-means for no. 4), and in some further 
cases experience
and research suggest typical behavior ($K$-means tends to produce clusters of
roughly equal size, whereas methods looking for high-density areas may produce
clusters of very variable size). See Section \ref{ch6.6:methods} for more
comments on specific methods. Other characteristics such as stability are
not involved in the definition of most clustering methods, but can be used to
validate clusterings and to compare clusterings from different methods.

The task of choosing a clustering method is made harder by the fact that in
many applications more than one of the listed characteristics is relevant.
Clusterings may be used for several purposes, or desired characteristics may 
not be well defined, e.g.,  in exploratory data analysis, or for realist 
clustering aims in cases where the connection between the interpretation of
the clusters and the data is rather loose. Also, a misguided 
desire for uniqueness and objectivity makes many researchers reluctant to
specify desired characteristics and choose a clustering method accordingly,
because they hope that there is a universally optimal method that will just
produce ``natural'' clusters. Probably for such reasons there is currently
almost no systematic research investigating the characteristics of
methods in terms of the various cluster characteristics. 

% {\bf Insert illustrating example if space}

\section{Data processing} \label{ch6.6:data}
The decision about what data to use, including how to choose, transform and 
standardize variables, and if and how to compute a dissimilarity measure, is
an important part of the methodological strategy in cluster analysis.
It often has a major impact on the clustering result, and is sometimes more
important than the choice of the clustering method.

\subsection{Choice of representation} \label{ch6.6:repr}
To some extent the data format restricts the choice of clustering methods; 
there are specialized methods for continuous, ordinal, categorical and mixed 
type data, dissimilarity data, graphs, time series, spatial data etc. But often
data can be represented in different ways. For example, a collection
of time series with 100 time points can be represented as points in 
100-dimensional Euclidean space, but they can also be represented by 
autocorrelation parameters of a time series model fitted to them, 
by wavelet features or some other low dimensional representation, or 
by dissimilarity measures which may involve some alignment or ``time warping'',
see \cite{Handbook}. On the other hand, dissimilarity data can be be 
transformed to Euclidean data using multidimensional scaling (MDS) 
techniques.
This means that the researcher often can choose whether the objects are
represented by features, dissimilarities, or in another way, for example by
vertices in a graph.

Generally, dissimilarity measures are a suitable basis for clustering if the
cluster concept is mainly based on the idea that similar objects should be 
grouped together and dissimilar objects should be in different clusters. 
Dissimilarity measures can be constructed for most data types.
On the other hand, clusters characterized by distributional and geometrical 
shapes and clusters with potentially high within-cluster variability or
skewness are found better with objects characterized by features instead of 
dissimilarities. 

The choice of representation should be guided by the question how
objects qualify to belong together in the same cluster. 
For example, if the data are time series, there are 
various different possible concepts of ``belonging together''. Time series
may belong together if their values are similar most of the time,
which is appropriate if the plain values play a large role in the assessment
of similarity (for example cigarettes smoked per day in research
about smoking behavior). A musical melody can be played at 
different speeds and in different keys, so that two musical melodies may still
be assessed as similar despite pitch values being quite different and changes 
in pitch happen at different times.  
In other applications, such as particle 
detection by electrodes, the characteristics of a single 
event that happens at a certain potentially flexible time point 
(such as a value 
going up and then down again) may be important, and having detected such an 
event, some specific characteristics of it may represent the objects in the most
useful manner. 

A central issue regarding the representation is the choice of variables that
are either used as features to represent the objects or on which a 
dissimilarity definition is based. Both subject matter and statistical 
considerations play a role here.  From a statistical point of view, a 
variable could be seen as 
uninformative if it is either strongly correlated with other variables and does 
not carry information on its own as long as certain other variables are kept,
or the variable may not be connected to any ``real'' clustering characterized
in the data for example by high density regions. Furthermore, in some 
situations (for example using gene expression data) the number of variables may
simply be so large that cluster analysis methods become unstable. 
There are various automatic methods for variable selection in connection with
clustering, see \cite{Handbook}
for clustering variables at the same time as observations, and 
\cite{AlTaLi14} for a recent survey. Popular
classical methods such as principal component analysis (PCA) and MDS 
are occasionally used for constructing informative variables. These, 
however, are based on objective functions (variance, stress) that do not 
have any relation to clustering, and may therefore miss information
that is important for clustering. There are some projection pursuit-type 
methods that aim at finding low-dimensional representations of the data 
particularly suitable for clustering (\cite{BolKrz03,TCDO09}).

It is important to realize, though, that the variables involved in clustering
define the meaning of the clusters. Changing the variables implies changing
the meaning. If the researchers have a clear idea about the meaning of the
clusters of interest, it is problematic to select variables in an automatic 
manner. For example, \cite{HenLi13} were interested in socio-economic
stratification, for which information on income, savings, education and
housing is essential. Even if for example incomes do not show any clear
grouping structure, or are correlated strongly with another variable, this 
does not constitute a valid reason to exclude this variable for constructing a 
clustering that is meant to reflect a meaningful socio-economic partition of
a population. A stratification based on automatically selected variables that
cluster in a nicer way may be of exploratory interest, but does not 
fulfill the aim of the researchers. One could argue that in case of correlation
between income and another variable, savings, say, 
the information from income is retained as long as savings (or a linear 
combination of them both, as would be generated by PCA) is 
still used as a feature for clustering. But this is not true, because the fact
that the information is shared by two variables that in terms of their meaning 
are essential for the clustering aim is additional information that should 
not be lost.

Another issue is that variables can play different roles, which has different
implications. For example, a dataset may include spatial coordinates and other
variables (e.g., regional data on avalanche risk, or color information in
image segmentation). Depending on the role that the spatial information should 
play, spatial coordinates can be included in the clustering process
as features together with the others (which implies that regional similarity 
will somehow be traded off against similarity regarding the other variables in
the clustering process), or they could define constraints (e.g.,
clusters on the other variables could be constrained to be spatially 
connected), or they could be ignored for clustering, but could be used 
afterward to validate the resulting clusters or to analyze their spatial
structure. For avalanche risk mapping, for example, one may take 
the latter approach for detailed maps if spatial information is discretized
and there is enough data at each point, but one may want to impose spatial 
constraints if data is sparser or if the map needs to be coarser because it is
used by decision makers instead of hikers.

Often there is a good reason for not choosing the variables automatically 
from the data, but rather guided by the aim of clustering. In some cases 
dimension reduction can be achieved by the definition of meaningful new 
indices summarizing information in certain variables. On the other hand, 
automatic variable selection may yield interesting clusterings if the aim 
is mainly exploratory, or if there is no prior information about
the importance of the variables 
and it is suspected that some of them are uninformative ``noise''.

\subsection{Dissimilarity definition} \label{ch6.6:diss}
In order to apply dissimilarity based methods and to measure whether
a clustering method groups similar observations together,
a formal definition of ``dissimilarity'' is needed 
(or ``proximity'', which refers to either dissimilarity or similarity,
as sometimes used in the literature; their treatment 
is equivalent and there are a number of transformations 
between dissimilarity and similarity measures, the simplest and most
popular of which
probably is ``dissimilarity$=$ maximum similarity minus similarity''). 
In many situations, dissimilarities between objects cannot be measured
directly, but have to be constructed from some measurements of variables of
the objects of interest. Directly measured dissimilarities occur
for example in comparative experiments in psychology and market research. 

There is no unique ``true'' dissimilarity measure for any dataset; the
dissimilarity measurement has to depend on the researchers' concept of what it
means to treat two objects as ``similar'', and therefore on the clustering aim. 

Mathematically, a dissimilarity is a function  
$d:\ {\cal X}^2\mapsto\R$, ${\cal X}$ being the object space, so that
$d(\mathbf{x},\mathbf{y})=d(\mathbf{y},\mathbf{x})\ge 0$ and 
$d(\mathbf{x},\mathbf{x})=0$ for $\mathbf{x},\mathbf{y}\in{\cal X}$.
There
is some work on asymmetric dissimilarities (\cite{Okada00}) and multiway
dissimilarities defined between more than two objects (\cite{Diatta04}). 
A dissimilarity
fulfilling the triangle equality 
$$d(\mathbf{x},\mathbf{y})+d(\mathbf{y},\mathbf{z})\ge 
d(\mathbf{x},\mathbf{z}),\ \mathbf{x},\mathbf{y},\mathbf{z}\in {\cal X}, $$
is called a ``distance'' or ``metric''. The triangle inequality is connected
to Euclidean intuition and therefore seems to be a ``natural'' requirement, 
but in some applications it is not appropriate. 
\cite{HenHau06} argue, e.g., that for presence-absence data of species
on regions two species A and B are very dissimilar if 
they are present on two small 
disjoint areas, but both should be treated as similar to a 
species C covering
a larger area that includes both A and B, if clusters are to be interpreted as
species grouped together by palaeoecological processes.

A vast number of dissimilarity measures has been proposed, some
for rather general purposes, some for more specific applications 
(dissimilarities between shapes (\cite{VelLat06}), melodies (\cite{MulFri07}),
geographical species distribution areas (\cite{HenHau06}), etc.). Chapter 3 in
\cite{EvLaLeSt11} gives a good overview of general purpose dissimilarities. 
Here are some basic considerations:
\begin{description}
\item[Aggregating binary variables.]  If two objects $\mathbf{x}_1, 
\mathbf{x}_2$ are represented by $p$ 
binary variables, let $a_{ij}$ be the number of variables $h=1,\ldots,p$ 
on which $x_{1h}=i, x_{2h}=j,\ i,j\in\{0,1\}$. If all variables are treated in
the same way, the most straightforward dissimilarity is the simple matching
coefficient, 
$$d_{SM}(\mathbf{x}_1,\mathbf{x}_2)=1-\frac{a_{00}+a_{11}}{p}.$$ 
However,
often (e.g. in the case of geographical presence-absence data in ecology)
common presences are important, whereas common absences are not. This is 
taken into account by the Jaccard dissimilarity 
$$d_J( \mathbf{x}_1,\mathbf{x}_2)=1-\frac{a_{11}}{a_{11}+a_{10}+a_{01}}.$$
One can worry about whether this gives the object with more presences
too much weight in the denominator, and actually more than 30 dissimilarity
measures for such data have been proposed \cite{Shi93}, prompting much
research about their characteristics and how they relate to each other 
(\cite{GowLeg86,War08}).
\item[Aggregating categorical variables.] If there are more than two 
categories, again the most intuitive way to construct a dissimilarity measure
is one minus the relative number of ``matches''. In some applications such 
as population genetics dissimilarity 
should rather be a non-linear function of matches between genes, and it is 
also important to think about whether and in what way variables with 
different numbers of categories or even with more or less uniform distributions
should be given different weights because some variables produce matches more
easily than others.
\item[Aggregating continuous variables.]
The Minkowski ($L_q$)-distance between
two objects $\mathbf{x}_i, \mathbf{x}_j$ on $p$ real-valued variables
$\mathbf{x}_i=(x_{i1},\ldots,x_{ip})$ is
\begin{equation}\label{ch6.6:eqminkowski}
  d_{Mq}(\mathbf{x}_i,\mathbf{x}_j)=\sqrt[q]{\sum_{l=1}^p d_l(x_{il},x_{jl})^q},
\end{equation}
where $d_l(x,y)=|x-y|$.  Variable weights $w_l$ can easily be incorporated by
multiplying the $d_l$ by $w_l$.
Most often, the Euclidean distance $d_{M2}$ and the Manhattan distance 
$d_{M1}$ are used. Using $d_{Mq}$ with larger $q$
gives the variables with larger $d_l$ more weight, i.e., two observations are
treated as less similar if there is a very large dissimilarity on one variable
and small dissimilarities on the others than if there is about the same 
medium-sized dissimilarity on all variables, whereas $d_{M1}$ gives all 
variable-wise contributions implicitly the same weight (note 
that this does not hold for the Euclidean distance that corresponds to physical 
distances and is used as default choice in many applications).

An alternative would be the (squared) Mahalanobis distance,
\begin{equation} \label{ch6.6:eqmahal}
  d_M(\mathbf{x}_i,\mathbf{x}_j)^2=(\mathbf{x}_{i}-\mathbf{x}_{j})^T{\bf S}^{-1}(\mathbf{x}_{i}-\mathbf{x}_{j}),
\end{equation}
where ${\bf S}$ is a scatter matrix such as the sample covariance matrix.
This is affine equivariant, i.e., not only rotating the data points in Euclidean
space, but also stretching them in any number of directions will not affect
the dissimilarity. It will also implicitly aggregate and therefore weight
information from strongly correlated variables down (correlation implies
that data are ``stretched'' in the direction of their dependence; the 
consequence is that ``joint information'' is only used once).
This is desirable if clusters can come in
in all kinds of elliptical shapes. On the other hand, it means that the weight
of the variables is determined by their covariance structure and not by their
meaning, which is not always appropriate (see the discussion about variable
selection above). 

There are many further ways of constructing a dissimilarity measure from
several continuous variables, see \cite{EvLaLeSt11}, 
such as the Canberra distance, which 
emphasizes differences close to zero. It is defined by
$q=1$ and $d_l(x,y)=\frac{|x-y|}{|x|+|y|}$ in (\ref{ch6.6:eqminkowski}).
The Pearson correlation coefficient $\rho(\mathbf{x},\mathbf{y})$ 
has been used to construct a dissimilarity measure 
$d_P(\mathbf{x},\mathbf{y})=1-\frac{\rho(\mathbf{x},\mathbf{y})+1}{2}$ as
well (other transformations are also used). This interprets $\mathbf{x}$ and
$\mathbf{y}$ as similar if they are positively linear dependent. This does 
not mean that their values have to be similar, but rather the values of the 
variables relative to the other variables. In some applications variables are
clustered, which means that variables and objects change their roles; if the 
variables are the objects to be clustered, $\rho$ in $d_P$ is a proper 
correlation between variables, which is a typical use of $d_P$.

\item[Aggregating ordinal variables.] Ordinal variables are characterized 
by the absence of metric information about the distances between two 
neighboring categories. They could be treated as categorical variables, but
this would ignore available information. On the other hand, it is fairly common 
practice to use plain Likert codes 1,2,\ldots and then to use methods for
continuous data. Ordinality can be taken into account while still using 
methods for continuous data by scoring the categories in a way that uses the
ordinal information only. Straightforward scores are obtained by ranking (using
the midrank for all objects in one category) or normal scores (\cite{Con99}),
which treat the data as if there would be an underlying uniform (ranks)
or Gaussian distribution (normal scores). A more
sophisticated approach is polytomous item response theory (\cite{OstNer06}). 
Using scores that are determined by the distribution of the data does
not guarantee that they appropriately quantify the interpretative distances
between categories, and in some situations (e.g., Likert scales in 
questionnaires where interviewees can see that responses are coded 1,2,\ldots)
this may be reflected better by plain Likert codes. Sometimes also there is
a more complex structure in the categories that can be reflected by scoring
data in a customized way. For example, in \cite{HenLi13}, a ``housing''
variable had levels ``owns'', ``pays rent'' and several levels such as ``shared
ownership'' that could be seen as lying between ``owns'' and ``pays rent'' but
could not be ordered, which could be reflected by having a distance of
1 between ``pays rent'' and ``owns'' and 0.5 between any other pair of 
categories.
\item[Aggregating mixed-type variables and missing values.]
If there are variable-wise distances $d_l$ defined, variables of mixed type
can be aggregated. A standard way of doing this is the Gower dissimilarity
(\cite{Gow71})
$$   d_{G}(\mathbf{x}_i,\mathbf{x}_j)=
\frac{\sum_{l=1}^p w_l \delta_{ijl} d_l(x_{il},x_{jl})}
{\sum_{l=1}^p w_l \delta_{ijl}},$$
where $w_l$ is a variable weight and $\delta_{ijl}=1$ except if 
$x_{il}$ or $x_{jl}$ are missing, in which case $\delta_{ijl}=0$. This is a 
weighted version of $d_{M1}$ and takes into account missing values by just
leaving the corresponding variable out and rescaling the others. 
Gower recommended to use the weight $w_l$ for standardization to $[0,1]$-range 
(see Section \ref{ch6.6stan}), but \cite{HenLi13} argued that many 
clustering methods tend to identify gaps in variable distributions with cluster
borders, and that this implies that $w_l$ should be used to weight 
binary and other ``very discrete'' variables down against continuous variables, 
because otherwise the former would get an unduly high influence on the 
clustering. $w_l$ can also be used to weight variables up that have
high subject matter importance.
The Gower dissimilarity is very general and covers most
applications of dissimilarity-based clustering to mixed-type 
variables. An alternative for missing values is to treat them as an own 
category. For continuous variables one could give missing values a
constant dissimilarity to every other value. More references are in 
\cite{EvLaLeSt11}.

\item[Custom-made dissimilarities for structured data.] In many situations 
detailed considerations regarding the subject matter will play the most 
important role regarding the design of a dissimilarity measure. This is 
particularly the case if the data are more structured than just a collection
of variables. Such considerations start with deciding how 
to represent the objects, as discussed in Section 
\ref{ch6.6:repr} and illustrated by the task of time series clustering.
The next task is how to aggregate the measurements in an appropriate way.
In time series clustering, one consideration is whether some processes that
are interpreted to be similar may occur at different and potentially
varying speeds, so that flexible alignment (``dynamic time warping'') is 
required, as may be the case in gesture recognition. See \cite{Liao05} for 
further aspects of choosing dissimilarities between time series.

Key issues may differ a lot from one application to the next, so it is 
difficult to present general rules. There is some research on approximating
expert judgments of similarity with functions of the available variables
(\cite{Gordon90,MulFri07}). \cite{HenHau06}, who incorporate geographical
distance information into a dissimilarity for presence-absence data,
list a number of general principles for designing and fine-tuning
dissimilarities:
\begin{itemize}
\item What should be the basic behavior of the dissimilarity as a function
of the existing measurements (when decreasing/increasing etc.)? 
\item What should be the relative weight of different aspects of the basic 
behavior? Should some aspects be incorporated 
in a nonlinear manner (see Section \ref{ch6.6:trans})?
\item Construct exemplary pairs of objects for which it is
clear what value the dissimilarity should have, or how it should
compare with some other exemplary pairs.
\item Construct sequences of pairs of objects in which one aspect changes
while others are held constant.
\item Whether and how could the dissimilarity measure be disturbed
by small changes in the characteristics? What behavior in these
situations would be appropriate?
\item Which transformations of
the variables should leave the dissimilarities unchanged?
\item Are there reasons that the dissimilarity measure should be a metric (or
have some other particular mathematical properties)?
\end{itemize}
\end{description}

\subsection{Transformation of variables}\label{ch6.6:trans}

According to the same philosophy as before, effective distances (as used by
a clustering method) on the variables should reflect the ``interpretative 
distance'' between objects, and transformations may be required to achieve
this. Because there is a large variety of clustering aims, it is difficult
to give general principles that can be applied in 
a straightforward manner, and the issue is best illustrated
using examples. Therefore, consider now the variable ``savings amount'' 
in socio-economic stratification in \cite{HenLi13}. Regarding
social stratification it makes sense to allow proportionally higher 
variation within high income and/or high savings clusters; the 
``interpretative difference'' between incomes is rather governed
by ratios than by absolute differences. In other words, the difference 
between two people with yearly incomes of \$ 2 million and \$ 4 million, say, 
should in terms of social strata be treated about equally as the
difference between \$ 20,000 and \$ 40,000. This suggests a log transformation,
which has the positive side effect to tame some outliers in the data.
Some people indeed have zero savings, which means that the transformation 
should actually be log(savings)$+c$. The choice of $c$ can have surprisingly
strong implications on clustering, because it tunes the size of the ``gap'' 
between persons with zero savings and persons with small savings; in the
dataset analyzed in \cite{HenLi13} there were only a handful of persons
with savings below \$ 100, but more with savings between \$ 100 and \$ 500. 
Clustering methods tend to identify borders between clusters with gaps. 
A low value for $c$, e.g., $c=1$, creates a rather broad gap, which means that
many clustering methods will isolate the zero savings-group regardless of the
values of the other variables. However, from the point of view of socio-economic
stratification, zero savings are not that special and not essentially different
from low savings below a few hundred dollars, and therefore a larger value
for $c$ (\cite{HenLi13} chose $c=50$) needs to be chosen to allow methods to 
put such observations together 
in the same cluster.  The reasoning may seem to be very subjective, but 
actually this is required when attention is paid to the detail, and there is
no better justification for any straightforward default choice (e.g., $c=1$).

It is fairly common that ``interpretative distances'' are nonlinear functions
of plain differences. As another example, \cite{HenHau06} used
geographical
distance information in a nonlinear way in a dissimilarity measure 
for presence-absence data for biological species, because individuals
can easily travel shorter distances, whereas what goes on in regions with
a long distance between them is rather unrelated, regardless of whether this
distance is, say, 2,000 or 4,000 km, the difference between which therefore
should rather be scaled down compared to differences between smaller distances.

Whether such transformations are needed depends on the clustering method.
For example, a typical distribution of savings amounts is very skew and 
sometimes the skewness corresponds to the change in interpretative distances
along the range of the variable.
Fitting a mixture of appropriate skew distributions (see \cite{Handbook})
can then have a similar effect as transforming the variable. 
   
\subsection{Standardization, weighting and sphering of variables}
\label{ch6.6stan}
Standardization of variables is a kind of transformation, but with a 
different rationale. Instead of governing the effective distance within 
a variable, it governs the relative weight of variables against each other when 
aggregating them. Standardization is not needed if a clustering method
or dissimilarity is used that is invariant against affine transformations
such as Gaussian mixture models allowing for flexible covariance matrices
or the Mahalanobis distance. Such methods standardize 
variables internally, and the following considerations may apply also to the
question whether it is a good idea to use such a method.

Standardization of $\mathbf{x}_1,\ldots,\mathbf{x}_n\in\R^p$
is a special case of the linear transformation
$$ \mathbf{x}^*_i=\mathbf{B}^{-1}(\mathbf{x}_i-\mathbf{\mu}),\ i=1,\ldots,n, $$
where $\mathbf{B}$ is an invertible 
$p \times p$-matrix and $\mathbf{\mu}\in \R^p$.
Standardizing location by
introducing $\mathbf{\mu}$ (usually chosen as the mean vector of the data) 
does not normally have an influence on clustering, but simplifies expressions.
``Standardization'' refers to using a diagonal matrix of scale statistics
(see below) as $\mathbf{B}$. For ``sphering'', $\mathbf{B}=\mathbf{UD}^{1/2}$, 
where  $\mathbf{S}=\mathbf{UDU'}$ for a scatter matrix $\mathbf{S}$, with 
$\mathbf{U}$ being the matrix of eigenvectors and $\mathbf{D}$ being the 
diagonal matrix of eigenvalues. 

If the clustering method is not affine invariant (for example 
$K$-means or dissimilarity-based methods using the Euclidean distance), 
standardization
may have a large impact. For example, if variables are measured on different 
scales and one variable has values around 1,000 and another one has values 
between 0 and 1, the first variable will dominate the clustering regardless
of what clustering pattern is supported by the second one. Standardization
makes clustering invariant against the scales of the variables, and sphering
makes clustering invariant against general affine linear transformations. 

But standardization and sphering are not always desirable. The effect of 
sphering is the same as the effect of using the Mahalanobis distance
(\ref{ch6.6:eqmahal}), discussed above. If variables
use the same measurement scale but have different variances, it depends on the
requirements of the application whether standardization is desirable or not.
For example, data may come from a questionnaire where respondents were asked
to rate several items on a scale between 1 and 10. If for some items almost all 
respondents picked central values between 4 and 7, this may well indicate that
the respondents did not find these items very interesting, and that therefore
these items are less informative for clustering compared with other items for 
which respondents made a good use of the full width of the scale. Fur standard
clustering methods that are not affine invariant, the variation within a 
variable defines its relative impact on the clustering. Leaving the
items unstandardized means that an item with little variation would have little
impact on clustering, which seems appropriate in this situation, whereas in
other applications one may want to allow the variables a standardized influence
on clustering regardless of the within-variable variation.

The most popular methods for standardization are
\begin{itemize}
\item standardization to $[0,1]$-range,
\item standardization to unit variance,
\item standardization to a unit value of a robust variance estimator such
as interquartile range (IQR) or median absolute deviation (MAD) from the 
median. 
\end{itemize}
As is the case for most such decisions, the standardization method 
occasionally makes a
substantial difference. The major 
difference is the treatment of outlying values. Range standardization is 
vulnerable to outlying values in the sense that an extreme outlier has the 
effect of squeezing together the other values on that variable, so that any
structural information in this variable apart from the outlier will only have
a very small influence on the clustering. This is avoided by using a robust
variance estimator, which can have another undesired effect. Although 
outliers on a single variable will not affect other structural information 
on the same variable so much, for objects for which a single variable has an
outlying value, this may dominate the information from all other variables, 
which can have a big impact in situations with many variables and a moderate
number of outlying values in various variables. Variance standardization 
compromises on the disadvantages of both other approaches as well as on the
advantages.

If for subject matter reasons some variables are more important than others
regardless of the within-variable variation,
one could reweight them by multiplying them with constants reflecting the
relative importance after having standardized their data-driven impact. 

None of the methods discussed up to here takes clustering information into 
account. A problem here is that if a variable shows a clear separation between
clusters, this may introduce large variability, which may imply a large
variance, range or IQR/MAD. If variables use the same measurement units and
values are comparable, this 
could be an argument against standardization; if within-cluster 
variation is low, range-standardization will normally be better than the 
other schemes (\cite{MilCoo88}). The problem is, obviously, that clustering 
information is not normally available a priori. \cite{ArGnKe82} discuss a
method in which there is an initial guess, based on smallest dissimilarities,
which objects belong to the same cluster, from which then a provisional
within-cluster covariance matrix is estimated, which is used to sphere the
dataset, \cite{DeSoete86} suggests to reweight variables in such a way that 
an ultrametric is optimally approximated (\cite{Handbook}). These
methods are compared with classical standardization by \cite{GnKeTs95}.

\section{Comparison of clustering methods}
\label{ch6.6:benchmark}
Different cluster analysis methods can be compared in several different ways. 
When choosing a method for a specific clustering aim, it is important to know
the characteristics of the clustering methods so that they can be 
matched with the required cluster concept. This is treated in Section
\ref{ch6.6:methods}. Section \ref{ch6.6:studies} 
reviews some existing studies comparing different clustering methods.
Section \ref{ch6.6:axioms} summarizes some theoretical work on desirable
properties of clustering methods.

\subsection{Relating methods to clustering aims}\label{ch6.6:methods}

Following Section \ref{ch6.6sec:philo}, the choice of an appropriate 
clustering method is strongly dependent on the aim of clustering.
Here I list some clustering methods treated in this book, and how they relate
to the list of potentially desirable cluster characteristics given in
Section \ref{ch6.6sec:philo}. Completeness cannot be achieved because of space
limitations. For definitions of all listed methods, see \cite{Handbook}.
\begin{description}
\item[$K$-means.] The objective function of $K$-means 
implies that it aims primarily at representing clusters by centroids. The
squared Euclidean distance penalizes large distances within clusters strongly,
so outliers can have a strong impact and there may be small outlying clusters,
although $K$-means generally rather tends to produce clusters of roughly equal
size. Distances in all directions from the center are treated in the same
way and therefore clusters tend to be spherical ($K$-means is equivalent to 
ML-estimation in a model where clusters are modeled by spherical
Gaussian distributions). $K$-means emphasizes homogeneity rather than 
separation; it is usually more successful regarding small within-cluster
dissimilarities than regarding finding gaps between clusters.  
\item[$K$-medoids] is similar to 
$K$-means, but it uses unsquared dissimilarities. This means that it may allow
larger dissimilarities within clusters and is somewhat
more flexible regarding outliers and deviations from the spherical cluster
shape. 
\item[Hierarchical methods.] A first consideration is 
whether a full hierarchy of clusters is required (for example because the
dissimilarity structure should be approximated by an ultrametric)
or whether using a 
hierarchical method is rather a tool to find a single partition by cutting
the hierarchy at some point. If only a single partition is required, 
hierarchies are not as flexible as some other algorithms for finding an
in some sense optimal clustering (this applies, e.g., to comparing Ward's
hierarchical method with good algorithms for the $K$-means objective function
as reviewed in \cite{Handbook}).
Different hierarchical methods produce quite different clusters. Both Single
and Complete Linkage are rather too extreme for many applications, although they
may be useful in a few specific cases. single 
linkage focuses totally on separation, i.e., keeping the closest points of
different clusters apart from each other, and Complete Linkage focuses totally
on keeping the largest dissimilarity within a cluster low. Most other
hierarchical methods are a compromise between these two extremes.
\item[Spectral clustering and graph theoretical methods.]
These methods are not governed by straightforward objective functions that
attempt to make within-cluster dissimilarities small or between-cluster 
dissimilarities large. Spectral clustering is connected to Single Linkage 
in the sense that its ``ideal'' clusters theoretically correspond to 
connected components of a graph. However, spectral clustering can be set up
in such a way (depending sometimes strongly 
on tuning decisions such as the how the edge 
weights are computed) that it works in a smoother and more flexible way than
Single Linkage, less vulnerable to single points ``chaining'' clusters. 
Generally spectral clustering still can produce very flexible cluster shapes
and focuses much more on cluster separation than on within-cluster homogeneity
when applied to originally Euclidean data in the usual way, i.e., using a 
strongly concave transformation of the dissimilarities so that the method
focuses on the smallest dissimilarities, i.e., the neighborhoods of
points, whereas pairs of points with large dissimilarity can still be connected
through chains of neighborhoods. 
\item[Mixture models.] The distributional
assumptions for such models define ``prototype clusters'', i.e., the 
characteristics of the clusters the methods will find. These characteristics
can depend strongly on details. For example, the Gaussian mixture model 
with fully flexible covariance matrices has a much larger flexibility 
(which often comes with stability issues and may incur quite large 
within-cluster dissimilarities) than a model in which covariance
matrices are assumed to be equal or spherical. Using mixtures of $t-$ or
very skew distributions will allow observations within clusters that are quite
far away from the cluster cores. Generally, the mixture model
does not come with implicit conditions that ensure the separation of clusters.
Two Gaussian distributions can be so close to each other that their mixture is
unimodal. Still, for a large enough dataset, the BIC will separate the two 
components, which is only beneficial if the clustering aim allows to split
up data subsets that seem rather homogeneous (the idea of merging such mixture
components is discussed in \cite{Handbook}). This issue is also important
to have in mind when fitting mixture models to structural data; slight 
violations of model assumptions such as linearity may lead to fits by more
``clusters'' that are not well separated, if the BIC is used to 
determine the number of mixture components. Standard latent class models
for categorical data assume local independence within clusters, which means
that clusters can be interpreted in terms of the marginal distributions of
the variables, which may be useful but is also restrictive, and allows 
large within-cluster dissimilarities. The comments here
apply for Bayesian approaches as well, which allow the user to ``tune'' the 
behavior of the methods through adjustment of the prior distribution, e.g., 
by penalizing methods with more clusters and parameters in a stronger
way. This can be a powerful tool for regularization, i.e., penalizing 
troublesome issues such as zero variances and spurious clusters.
On the other hand, such priors may have unwanted implications. For 
example, the Dirichlet prior implies that a certain non-uniform 
distribution of cluster sizes is supported.
\item[Clustering time series, functional data and symbolic data.] 
As was already discussed in Section 
\ref{ch6.6:repr}, regarding time series and also functions and symbolic data, 
a major issue to decide is in what sense the sequences of observations should
belong together in a cluster, which could mean for example 
similar values, similar 
functional shapes (with or without alignment or ``time warping''),  similar
autocorrelation structure, or good approximation by prototype objects. This is
what mainly distinguishes the many methods discussed in these chapters.
\item[Density-based methods.] Identifying
clusters with areas of high density seems to be very intuitive and directly 
connected to the term ``cluster''. High density areas can have very 
flexible shapes, but more sophisticated density-based methods do not depend as
strongly on one or a few points as Single Linkage, which can be seen as a 
density-based method. There are a few potential peculiarities 
to keep in mind. High density areas may vary a lot in size, so they 
may include very large dissimilarities and there may be much variation in 
numbers of points per cluster. In different locations in the same dataset,
depending on the local density, different density levels may qualify as 
``high'', and methods looking for high density areas at various resolutions 
can be useful. Clusters may also be identified with density modes, which occur
at potentially very different density levels. Density-based methods usually 
do not need the number of clusters specified, but rather their resolution, 
i.e., size of neighborhood (in terms of number of neighbors or radius), grid 
size or kernel bandwith. This determines how large gaps in the density have 
to be in order to be found as cluster borders and is often not easier than
specifying the number of clusters. In higher dimensions, it becomes more 
difficult for clustering algorithms to figure out properly where the density
is high or low, and also the sparsity of data in high-dimensional space means 
that densities tend to be more uniformly low.
\end{description}
% For choosing a clustering method, it is then necessary to know how they 
% correspond to the required characteristics. Some methods optimize certain
% characteristics directly (such as $K$-means or PAM for no. 4, 
% fitting mixture distributions for no. 3, and latent class analysis for
% no. 11), and in some further cases experience
% and research suggest typical behavior ($K$-means tends to produce clusters of
% roughly equal size, whereas methods looking for high-density areas may produce
% clusters of very variable size). Other characteristics such as stability are
% not involved in the definitions of clustering methods, but can be used to
% validate clusterings and to compare clusterings from different methods.
% 
% 
% Also distance transformation in spectral clustering
% 

\subsection{Benchmarking studies}
\label{ch6.6:studies}
Different clustering methods can be compared based on datasets in which a 
true clustering is known. There are three basic approaches for this in the
literature (see \cite{Hen15} for more discussion and some philosophical 
background regarding the problem of defining the ``true'' clusters):
\begin{enumerate}
\item Real datasets can be used in which there are known classes of some kind
(a problem with this is that there is no guarantee that the known ``true'' 
classes are the only ones that make sense, or that they even cluster properly
at all).
\item Data can be simulated from mixture or fixed partition models where 
within-cluster distributions are homogeneous, such as the
Gaussian or uniform distribution (it depends on the
separation of the mixture components whether these can be seen as separated 
clusters; also such datasets will naturally favor clustering methods that are
based on the corresponding model assumptions).
\item Real data can be used for which there is no knowledge of a true 
clustering. 
\end{enumerate}
Measures as introduced in \cite{Handbook} such as the adjusted Rand index
can then be used in order to 
compare the results of clustering methods with the true clusterings in the
first two approaches. Measuring the quality of the clusterings for the third
approach is less straightforward, and this is used less often. 
\cite{MoBlSk83}, for example, used a dataset of 750 alcohol abusers on
some socio-behavioral variables, and measured quality by external validation,
i.e. looking at the discrimination of the clusters by some external variables,
and by splitting the data into two random subsamples, clustering both,
and using nearest centroid allocation for computing a similarity measure of
the clustering of the different subsamples. Another approach is to compare 
dissimilarity data to the ultrametric induced by a hierarchical clustering
using the cophenetic correlation, see \cite{Handbook}, as done by 
\cite{SaDoDo13} for artificial data.

At first sight it seems to be a very important and promising project to 
compare clustering methods comprehensively, given the variety of existing 
approaches that is often confusing for the user. Unfortunately, the variety of 
clustering aims and cluster concepts and also the variety of possible datasets,
both regarding data analytic features such as shape of clusters, number of
clusters, separation of clusters, outliers, noise variables, 
and regarding data formats (Euclidean, 
ordinal, categorical variables, number of variables, structured data, 
dissimilarity data of various different kinds) makes such a project a 
rather unrealistic prospect. 

In the 1970s and 1980s, with less methodology already existing, 
a number of comparative
benchmark studies were run on artificial data, usually focusing on 
standard hierarchical methods
and different $K$-means-type algorithms. Some of these (the most comprehensive 
of which was \cite{Mil80}) are summarized in \cite{Mil96}. As could be 
expected, results depended heavily on the features of the datasets. Overall, 
Ward's hierarchical clustering seemed rather successful and single linkage 
seemed problematic, although at least the first result may be
biased to some extent by the data generation processes used in
these studies.
 
More recent studies tend to focus on more specialist issues such as comparing
different algorithms for the $K$-means criterion (\cite{BruSte07}),
comparing $K$-means with Gaussian mixture models with more general covariance
matrix models (\cite{SteBru11}; note that the authors show that often
$K$-means does rather well even for non-spherical data, but this work is 
a discussion paper and some discussants highlight situations where
this is not the case), or a latent class mixture model and $K$-medoids for
categorical data (\cite{AndHen14}). \cite{DiBaWiHoMo04} is an example for
a study on data typical for a specific application, namely 
functional magnetic resonance imaging datasets. The winners of their study 
are neural gas and $K$-means.

A large number of comparative 
simulation studies can be found in papers that introduce 
new clustering methods. However, such studies are usually often biased in 
favor of the new method that the author wants to advertise by showing that 
it is superior to some existing methods. Although such studies potentially 
contain interesting information about how clustering methods compare, having
their huge number and strongly varying quality in mind, the author takes 
the freedom to cite as a single example
\cite{CorHen14}, comparing robust clustering methods
on Euclidean data with elliptical clusters and outliers.

A very original approach was taken by \cite{JTLB04}, who did not attempt
to rank clustering methods according to their quality. Instead, they
clustered 35 different clustering algorithms into five 
groups based on their partitions of twelve different datasets. 
The similarity between the clustering
algorithms was measured as the averaged similarity (Rand index) between
the partitions obtained on the datasets. Given that different clustering
methods serve different aims and may well arrive at different legitimate 
clusterings on the same data, this seems to be a very appropriate approach. 
Apart from already mentioned methods, this study includes a number of graph
based and spectral clustering algorithms, some methods optimizing 
objective functions other than $K$-means (CLUTO), and ``Chameleon-type''
methods, i.e. more recent hierarchical algorithms based on dynamic modeling.

Still, it is fair to say that existing work merely scratches the surface of 
what could be of potential interest in cluster benchmarking, and there is
much potential for more systematic comparison of clustering methods. 

\subsection{Axioms and theoretical characteristics of clustering methods}
\label{ch6.6:axioms}
Another line of research aims at exploring whether clustering methods
fulfill some theoretical desiderata. 
\cite{JarSib71} listed a number of supposedly 
``natural'' axioms for clustering methods and showed that single linkage
was the only clustering method fulfilling them. Single Linkage also 
fulfills eight out of 
nine of the admissibility criteria given in \cite{FiVN71}, more than any 
other method compared there (which include standard hierarchical methods and 
$K$-means). Together with the fact that Single Linkage is known to be 
problematic in many situations because of chaining phenomena and the 
possibility to produce very large within-cluster dissimilarities, these results
should indeed rather put into question the axiomatic approach than all 
methods other than single linkage. Both these papers motivate their axioms 
from intuitive considerations, which can be criticized (see, e.g., 
\cite{KaRo90}). 
It turns out that monotonicity axioms are among the most 
restrictive. \cite{JarSib71} discuss clustering methods that map 
dissimilarities $d$ to clusterings that can be represented by ultrametrics
$u=F(d)$, such as most standard hierarchical clustering methods, and their 
monotonicity axiom requires $d\le d'\Rightarrow F(d)\le F(d')$. From the point
of view of ultrametric representation of a distance this may look harmless,
but in fact the axiom   
restricts the options for partitioning the data at the different levels
of the hierarchy quite severely, because it implies that if $d(a,b)$ is 
increased for two
observations $a$ and $b$ that are in the same cluster at some level, 
neither $a$ nor $b$ nor other points in this cluster can be merged with 
points in other clusters on a lower level as a result of the modification.

\cite{FiVN71} use a variant of this criterion, which 
requires that the resulting clustering does not change, and is therefore 
applicable to procedures that do not yield ultrametrics. The implications are
similarly restrictive. They state explicitly that some admissibility criteria
only make sense in certain applications. For example, they define ``convex
admissibility'', which states that the convex hulls of different clusters do not
intersect. This requires the data to come from a linear space and rules out
certain arrangements of nonlinear shaped clusters. It is the only 
criterion in \cite{FiVN71} that is violated by single linkage. Other 
admissibility 
criteria are concerned with a method's ability to recover certain ``strong''
clusterings, e.g., where all within-cluster dissimilarities are smaller than 
all between-cluster dissimilarities.

More recently, there is some revived interest in the axiomatic
characterization of clustering methods. \cite{Kle02} 
proved an ``impossibility theorem'', stating that there
can be no partitioning method fulfilling a set of three conditions claimed to
be ``natural'', namely
scale invariance (multiplying all dissimilarities with a constant does not
change the partition), richness (any partition of points is a possible outcome
of the method; this particularly implies that the number of clusters cannot 
be fixed) and consistency. The latter condition states that if the 
dissimilarities are changed in such a way that all 
within-cluster dissimilarities are made smaller or equal, and all 
between-cluster dissimilarities are made larger or equal, the clustering
remains the same. Like the monotonicity axioms before, this is more 
restrictive than the author suggests, because the required transformation
can be defined in such a way that two or more very homogeneous subsets
emerge within a single original cluster, which intuitively suggests that the
original cluster should then be split up (a corresponding
relaxation of the consistency condition is proposed in the paper and does not
lead to an impossibility theorem anymore). Furthermore, \cite{Kle02} shows that
three different versions of deciding where to cut a Single
Linkage dendrogram can fulfill any two of the three conditions, which means 
that these conditions cannot be used to distinguish any other
clustering approach from single linkage. 

\cite{AckBD08} respond to
Kleinberg's paper. Instead of using the axioms to characterize clusterings,
they suggest to use them (plus some others) 
to characterize cluster quality functions (CQF), 
and then clusterings could be found by optimizing these functions. Note that a 
clustering method optimizing a consistent CQF (i.e., a CQF that cannot become
worse under the kind of transformation of dissimilarities explained above)
does not necessarily yield consistent clusterings, because in a modified 
dataset other clusterings could look even better. The idea also applies with
modified axioms to clustering methods with fixed number of clusters.
Follow-up work studies specific properties of clustering methods with the aim
of providing axioms that serve to distinguish clustering methods as suitable
for different applications (\cite{AckBDLok10,ABDBL12}).
A similar approach is taken by \cite{PuHoBu00}, who compare a number of 
clustering criteria based on separability measures
averaging between-cluster dissimilarities in different ways according to a set
of axioms some of which are very similar to the above, adding local shift
invariance and robustness criteria that formalize that small changes to 
single dissimilarities can only have limited influence on the criterion.

\cite{CorMor13} starts from \cite{Kle02} in a different way and allows 
clustering methods to be restricted by certain parameters (such as the number
of clusters). The axioms apply to clusterings as in \cite{Kle02}, but a
number of variants of the consistency requirement are defined, and several
clustering methods including Single and Complete Linkage and $K$-means are
shown to be scale invariant, rich and consistent in a slightly re-defined 
sense.

Still, much existing work on axiomatic characterization is concerned 
with distinguishing ``admissible'' from ``inadmissible'' methods, exceptions
being \cite{AckBDLok10,ABDBL12}. This is of
limited value in practice, particularly because up to now no method in at least
fairly widespread use has been
discredited because of being ``inadmissible'' in such a theoretical sense;
in case of negative results, rather the admissibility criteria were put into 
question. Still
there is some potential in such research to learn about the 
clustering methods. Changing the focus from branding methods as generally
inadmissible to distinguishing the merits of different approaches seems to be
a more promising research direction.
A number of other characteristics of clustering methods has been studied
theoretically, see for example the references on robustness and stability
measurement in \cite{Handbook}. 

\cite{AckBD09} axiomatize ``clusterability'' of datasets with a view towards
finding computationally simpler algorithms for datasets that are ``easy''
to cluster, which mainly means that there is strong separation between
the clusters.

% \section{Cluster validation and comparing clusterings}
\section{Cluster validation}
\label{ch6.6:validation}
Cluster validation is about assessing the quality of a clustering on a dataset
of interest. Different from Section \ref{ch6.6:studies}, here the focus is 
on analyzing a real dataset for which the
clustering is of real interest, and where no ``true'' clustering is known with
which the clustering to be assessed could be compared (the approaches 
in Sections \ref{ch6.6:external}, \ref{ch6.6:index} and \ref{ch6.6:stability}
can also be used in benchmarking studies). 
Quality assessment of a single clustering
can be of interest in its own right, but methods for assessing the cluster 
quality can also be used for comparing different clusterings, be they from 
different methods, or from the same method but with different input parameters,
particularly with different numbers of clusters. Because the latter is a 
central problem in cluster analysis, some literature uses the term ``cluster 
validation'' exclusively for methods to decide about the number of clusters, 
but here a more general meaning is intended.

In any case cluster validation is an essential step in the cluster analysis
process, particularly because most methods do not come with any indication
of the quality of the resulting clustering other than the value of the 
objective function to be optimized, if there is one. 

% \subsection{Assessing the quality of a clustering}
There are several different approaches to cluster validation. \cite{Hen05}
lists
\begin{itemize}
\item use of external information,
\item testing for clustering structure,
\item internal validation indices,
\item stability assessments,
\item visual exploration,
\item comparison of several different clusterings on the same dataset.
\end{itemize}
Before going through these, I start with some considerations regarding the 
decision about the number of clusters.
\subsection{The number of clusters}
\label{ch6.6noc}
As the clustering problem as a whole, also the problem of deciding
the number of clusters is not uniquely defined, and there is 
no unique ``true'' number of clusters. Even if the clustering method is 
chosen, the number of clusters is still ambiguous. The ideal situation for
defining the problem properly seems to be if data are assumed to come from
a mixture probability model, e.g., a mixture of Gaussians, and every mixture 
component is identified with a cluster. The problem then seems to boil down
to estimating the number of mixture components. To do this consistently 
is difficult enough
(see \cite{Handbook}), but unfortunately in reality it is an ill-posed
problem. Generally, probability models are not expected to hold precisely
in reality. But if the data come from a distribution that is not exactly a 
Gaussian mixture with finitely many components, a consistent criterion (such
as the BIC, see \cite{Handbook}) will estimate a number of clusters 
converging to infinity, because a large dataset can be approximated better with
more mixture components. If mixture components are to be interpreted as
clusters, normally at least some separation between them is required, which
is not guaranteed if their number is estimated consistently.

The decision about which number of clusters is appropriate in 
a certain application amounts to deciding in some way what granularity is
required for the clustering. Ultimately, how strong separation between 
different clusters is required and a partition into how many clusters is useful
in the given situation cannot be decided by the data alone without user input. 
It is often suggested in the literature that the number of clusters 
needs to be ``known'' or otherwise it needs to be estimated from the data.
But if it is understood that finding the number of clusters in a 
certain application needs user input anyway, fixing the number of clusters
is often as legitimate a user decision as the user input needed otherwise. 
There are many
supposedly ``objective'' criteria for finding the best number of clusters
(see \cite{Handbook}). But
it would be more appropriate to say that these criteria, instead of estimating
any underlying ``true'' number of clusters, implicitly {\it define} what the
best number of clusters is, and the user still needs to decide
which definition is appropriate in the given application. 

In many
situations there are good reasons not to fix the number of clusters but rather
to give the data the chance to pick a number that fits its pattern. But the
researcher should not be under the illusion that this can be done reliably 
without having thought thoroughly about what cluster concept is required.
Apart from the indices listed in \cite{Handbook}, also the statistics 
listed in Section \ref{ch6.6:index} can be used, particularly if the researcher
has a quantitative idea about, for example, how strong separation between
clusters is required.

\subsection{Use of external information} \label{ch6.6:external}
Formal and informal external information
can be used. Informally, subject matter experts can often decide to what 
extent a clustering makes sense to them. On one hand, this is certainly not 
totally reliable, and a clustering that looks surprising to a subject matter
expert may even be particularly interesting and could spark new discoveries.
On the other hand, the subject matter expert may have good reasons to discard
a certain clustering, which often points to the fact that the clustering aim
was not well enough specified or understood when choosing a certain clustering
method in the first place. If possible, the problem should then be understood 
in such a way that it can lead to an amendment in the choice of methodology.

For formal external validation, there may be external variables or groupings 
known that are expected or desired to be related to the clustering. For 
example, in market segmentation, a clustering may be computed of data that 
gives preferences of customers for certain products or brands, and in order to
make use of these clusters, they should be to some extent homogeneous also 
regarding other features of the customers such as sex, age, household size
etc. This can be explored using techniques such as MANOVA and discriminant
analysis for continuous variables, and association measures or tests and
measures for comparing clusterings (see \cite{Handbook}) for categorical
variables and groupings.
    
\subsection{Testing for clustering structure} \label{ch6.6:test}
In many clustering applications, researchers may want to determine whether 
there is a ``real'' clustering in the data that corresponds to an underlying
meaningful grouping. Many clustering algorithms deliver a clustering regardless
of whether the dataset is ``really'' clustered. A chapter in \cite{Handbook} 
is about 
methods to test homogeneity models against  
clustering alternatives. Note that straightforward models for homogeneity such 
as the Gaussian or uniform distribution may be too simple to model even some
datasets without meaningful clusters. Significant deviations from such 
homogeneity models may sometimes be due to outliers, skew or 
nonlinear distributional shapes, or other structure in the data such as 
temporal or spatial autocorrelation, in which case it is advisable to use
more complex null models, see \cite{Handbook}. In any case it is 
important that a significant result of a homogeneity test does not necessarily
validate every single one of the found clusters. Homogeneity tests have 
been applied to single clusters or pairs of clusters in order to give more local
information about grouping structure, but this is not without problems, see \cite{Handbook}.

\subsection{Internal validation indices} \label{ch6.6:index}
A large number of indices has been proposed in the literature for 
evaluating the quality of a clustering based on the clustered data alone.
Such indices are comprehensively discussed in \cite{Handbook}. Most of
them attempt to summarize the clustering quality as a single number, which
is somewhat unsatisfactory according to the discussion in Section 
\ref{ch6.6sec:philo}.

Alternatively it is possible to measure relevant aspects of a clustering
separately in order to characterize the cluster quality in a multivariate way.
Indices measuring several aspects of a clustering are
implemented in the R-package ``fpc''. Here are some examples:
\begin{itemize}
\item measurements of within-cluster homogeneity such as maximum or average 
within-cluster dissimilarity, within-cluster sum of squares, or the largest
within-cluster gap;
\item measurements of cluster separation such as the minimum or average
dissimilarity between clusters; \cite{Hen14} proposes 
the average minimum dissimilarity to a point from a different cluster 
of the 10\% of observations for which this is smallest;
\item measurements of fit such as within-cluster sum of dissimilarities
from the centroid or Hubert's $\Gamma$-type measures, see \cite{Handbook};
\item measurements of homogeneity of different clusters, e.g., the entropy
of the cluster sizes or the coefficient of variation of cluster-wise average 
distances to the nearest neighbor; 
\item measurements of similarity between the empirical within-cluster
distribution and distributional shapes of interest, such as the Gaussian or
uniform distribution.
\end{itemize}

\subsection{Stability assessment}
\label{ch6.6:stability}
Stability is an important aspect of clustering quality. Certainly a clustering
does not warrant a strong interpretation if it changes strongly under slight
changes of the data. Although there is theoretical work on clustering stability
(see \cite{Handbook}), this gives very limited information about
to what extent a specific clustering on a specific dataset is stable.

Given a dataset, stability can be explored by generating artificial variants
of the data and exploring how much the clustering changes. This is treated 
in \cite{Handbook}. Standard
resampling approaches are nonparametric bootstrap, subsampling and 
splitting of the dataset. Alternatively, observations may be ``jittered'' or
additional observations such as outliers added, although the latter 
approaches require a model for adding or changing observations. 

Aspects to keep in mind are firstly that often parts of the dataset are clearly
clustered and other parts are not, and therefore it may happen that some
clusters of a clustering are stable and other parts are not. Secondly,
stability is not enough to ensure the quality or meaningfulness
of a clustering. For example, 
a big enough dataset from a homogeneous distribution may allow a very stable
clustering. For example, 2-means will partition data from a uniform distribution
on a two-dimensional rectangle in which one side is twice as long as the other
in a very stable manner with only a few ambiguities along the borderline 
of the two clusters. Thirdly, in some applications in which data are clustered
for organizational reasons such as information reduction, 
stability is not of much interest.

\subsection{Visual exploration}
\label{ch6.6:visual}
The term ``cluster'' has an intuitive visual meaning to most people, and also
in the literature about cluster analysis visual displays are a major device
to introduce and illustrate the clustering problem. Many of the potentially
desired features of clusterings such as separation between clusters, 
high density within clusters, and distributional shapes
can be explored graphically in a more holistic (if subjective) way than by
looking at index values. Standard visualization
techniques such as scatterplots, heatplots and mosaic plots for categorical
data as well as interactive and dynamic graphics 
can be used both to find and to validate clusters, see, e.g, \cite{TheUrb08}, 
\cite{CooSwa07}. For cluster validation, one would normally distinguish the
clusters using different 
colors and glyphs. Most people's intuition for 
clusters is strongly connected to the low-dimensional Euclidean space, and 
therefore methods that project data into a low-dimensional Euclidean space such
as PCA are
popular and useful. A chapter in \cite{Handbook} illustrates the use of PCA and a number of other techniques for
cluster visualization with a focus on network-based techniques and visualization
of curve clustering. There are also specialized 
projection techniques for visualizing the
separation between clusters in a given clustering 
(\cite{Hen04}) and for finding clusters (\cite{BolKrz03,TCDO09}). \cite{Hen05}
proposes to look for every single cluster at plots that show its separation
from the remainder of the dataset, as well as projection pursuit plots for the
data of a single cluster on its own to detect deviations from homogeneity.
Such plots can also be applied to more general data formats if a dissimilarity
measure exists by use of MDS. The implementation
of MDS in the ``GGvis'' package allows dynamic and interactive 
exploration of the data and of the parameters of the MDS 
(\cite{Buja08datavisualization}). \cite{AndHen14} apply MDS to visualize 
clusters in categorical data. 

A number of visualization methods have been developed specifically for 
clustering, of which dendrograms (see \cite{Handbook}) are probably 
most widespread. Dendrograms are also frequently used for ordering observations
in heatplots. Due to their ability to visualize high-dimensional information
and dissimilarity matrices without projecting on a lower-dimensional space,
heatplots are often used for such data. Their use depends heavily on the order
of the observations. For use in cluster validation it is desirable to plot
observations in the same cluster together, which is achieved by the use of
dendrograms for ordering the observations. However, it would also be desirable
to order observations within clusters in such a way that the transition between
clusters is as smooth as possible, so that not well separated clusters can be
detected. This is treated by \cite{HahHor11}.

\cite{KaRo90} introduced the silhouette plot based on the silhouette width
(see \cite{Handbook}), which shows how well observations are 
separated from neighboring clusters. In \cite{Joe04} this is compared with
plots based on the within-cluster data depth. \cite{Leisch10} introduces
another alternative to the silhouette width based on centroids along with 
further plots to explore how clusters are concentrated around cluster
centroids.

\subsection{Different clusterings on the same dataset}
\label{ch6.6:compclu}
The similarity between different clusterings on the same dataset can be
measured using the ideas in the corresponding chapter of \cite{Handbook}. 
Running different cluster
analyses on the same dataset and analyzing to what extent the results 
differ can be seen as an alternative approach to find out whether and which
clusters in the dataset are stable and meaningful. Some care is required
regarding the choice of clustering methods and the interpretation of results.
If certain characteristics of a clustering are important in a certain 
application and others are not, it is more important that the chosen cluster
analysis method delivers a good result in this respect than that its results 
coincide largely with the results of a less appropriate method. So if 
methods are chosen that are too different from each other, some of them may 
just be inappropriate for the given problem and no importance should be 
attached to their results. On the other hand, if too similar methods are chosen
(such as Ward's method and $K$-means), the fact that clusters are similar 
does not tell the user too much about their quality. Looking at the similarity
of different clusterings on the same data is useful mainly for two reasons:
\begin{itemize}
\item Several different methods may seem appropriate for the clustering aim,
either because the aim is imprecise, or because heterogeneous and 
potentially conflicting characteristics of the clustering are desired.
\item Some fine-tuning is required (such as neighborhood sizes in density-based
clustering, variable weighting in the dissimilarity, or prior specifications in
Bayesian clustering), and it is of interest
to explore how sensitive the clustering solution is to such tuning, 
particularly because the precise values of tuning constants are hardly 
fully determined by background knowledge.
\end{itemize}

\section{Conclusions}\label{ch6.6:conclusion}
In this paper, the decisions required for carrying out a cluster analysis 
are discussed, connecting them 
closely to the clustering aims in a specific application. The paper
is intended to serve as a general guideline for clustering and for choosing
the appropriate methodology from the many approaches on offer in 
\cite{Handbook}. 
~\\~\\
{\bf Acknowledgement:} This work was supported by EPSRC grant EP/K033972/1.

\bibliographystyle{chicago}

\bibliography{biblist6.6arxiv}

\end{document}